\pdfoutput=1

\documentclass[aps,amsmath,twocolumn, amssymb,preprintnumbers,superscriptaddress,floatfix]{revtex4-2}

\usepackage[utf8]{inputenc}
\usepackage{newtxtext}
\usepackage[upint]{newtxmath}
\usepackage{microtype}
\usepackage{textcomp}
\usepackage{eucal}
\usepackage{bm}

\usepackage{enumerate}
\usepackage{amsfonts}
\usepackage{amsmath}
\usepackage{amssymb}
\usepackage{color}
\usepackage{soul}

\usepackage{graphicx}

\usepackage[colorlinks,allcolors=blue]{hyperref}
\usepackage[capitalize]{cleveref}

\let\propto=\sim
\let\phi=\varphi
\let\epsilon=\varepsilon

\newcommand{\e}[1]{\mathrm{e}^{#1}}

\newcommand{\para}{{\parallel}}

\newcommand{\vecq}{{\boldsymbol{q}}}

\newcommand{\vecz}{{\boldsymbol{z}}}

\newcommand{\vecp}{{\boldsymbol{p}}}
\newcommand{\veck}{{\boldsymbol{k}}}

\newcommand{\vecA}{{\boldsymbol{A}}}

\newcommand{\vecM}{{\boldsymbol{M}}}

\newcommand{\vecR}{{\boldsymbol{R}}}

\newcommand{\vecsigma}{{\boldsymbol{\sigma}}}
\newcommand{\vecn}{{\boldsymbol{n}}}

\newcommand{\cg}{\check{g}}
\newcommand{\bha}{\hat{\boldsymbol{A}}}
\newcommand{\tnabla}{\tilde{\nabla}}

\newcommand{\etal}{et al.\ }
\newcommand{\eg}{\textit{e.g.}\ }

\newcommand{\Tr}[0]{\text{Tr}}
\renewcommand{\vec}[1]{\bm{#1}}
\newcommand{\pv}[0]{\bm{p}}
\newcommand{\rv}[0]{\bm{r}}

\newcommand{\qv}[0]{\bm{q}}

\newcommand{\Tm}[0]{\mathcal{T}}

\definecolor{DarkGray}{rgb}{0.7,0.7,0.7}

\newcommand{\prlsection}[1]{\textit{#1}.\kern0.05em---\kern0.05em\ignorespaces}

\begin{document}
\title{Quasiclassical boundary conditions for spin-orbit coupled interfaces with spin-charge conversion}

\author{Jacob Linder}
\thanks{Both authors contributed equally to this work.}
\affiliation{Center for Quantum Spintronics, Department of Physics, Norwegian \\ University of Science and Technology, NO-7491 Trondheim, Norway}
\author{Morten Amundsen}
\thanks{Both authors contributed equally to this work.}
\affiliation{Nordita, KTH Royal Institute of Technology and Stockholm University,
Hannes Alfvens vag 12, SE-106 91 Stockholm, Sweden}

\date{\today}
\begin{abstract}
The quasiclassical theory of superconductivity provides a methodology to study emergent phenomena in hybrid structures comprised of superconductors interfaced with other materials. A key component in this theory is the boundary condition that the Green functions describing the materials must satisfy. Recently, progress has been made toward formulating such a boundary condition for interfaces with spin-orbit coupling, the latter playing an important role for several phenomena in spintronics. Here, we derive a boundary condition for spin-orbit coupled interfaces that includes gradient terms which enables the description of spin-Hall like effects with superconductors due to such interfaces. As an example, we show that the boundary conditions predict that a supercurrent flowing through a superconductor that is coupled to a normal metal via a spin-orbit interface can induce a non-local magnetization in the normal metal.
\end{abstract}

\maketitle

\section{Introduction}
Superconductors \cite{bcs} carry charge currents without resistance in their natural state, but can also be made to transport spin currents without resistance. This happens when the spinless Cooper pairs in a superconductor are transformed to so-called triplet Cooper pairs \cite{maeno_rmp_03} that carry spin. Such a transformation has been predicted \cite{bergeret_rmp_05, buzdin_rmp_05} to take place in two ways: either by placing the superconductor in contact with a ferromagnetic material or by placing it in contact with spin-orbit coupled materials or impurities \cite{bergeret_prb_14}. Besides making the superconductivity spin-polarized \cite{linder_nphys_15, eschrig_rpp_15}, such hybrid structures also give rise to odd-frequency superconductivity \cite{berezinskii, linder_rmp_19} where electrons are non-locally correlated in time. Although transport of charge is already dissipationless in superconductors, it is both of fundamental and practical interest to study dissipationless transport of spin via superconductors~\cite{Ohnishi2020}. Indeed, lossless spin transport can offer new types of functionality in solid-state devices when combining superconductors with magnetic materials that charge-based transport is not capable of. Examples of this includes magnetization switching and domain wall/skyrmion motion with very low dissipation of energy \cite{brataas_natmat_12}. In addition, triplet superconductivity can provide novel means of probing quantum materials~\cite{Han2020}, as well as giving rise to functional material properties that are unique to the superconducting state, without any counterpart in conventional spintronics---such as quantum phase batteries \cite{szombati_nphys_15}. \\

Several experiments have confirmed theoretical predictions in the case of hybrid structures comprised of superconductors and ferromagnets \cite{ryazanov_prl_00, kontos_prl_01, keizer_nature_06, khaire_prl_10, robinson_science_10, kolenda_prl_16, birge_natphys_16}. On the other hand, experimental studies of triplet superconductivity in hybrid structures of superconductors and spin-orbit coupled materials have started to appear mostly in recent years \cite{banerjee_prb_18, satchell_prb_18, satchell_prb_19, martinez_prapp_20, gonzalez-ruano_prb_20}. There exists several predictions \cite{reeg_prb_15, hogl_prl_15, bobkova_prb_17, mishra_prb_21, costa_prb_21} on emergent phenomena in structures combining superconducting and spin-orbit coupled materials which remain to be tested experimentally, where the main theme is that spin-orbit interactions exists in a material due to the lack of inversion symmetry or due to impurities. 

Spin-orbit coupling can be amplified at interfaces involving heavy metals such as Pt or W. Such interfaces are known to play an important role in conventional spintronics \cite{amin_jap_20}. Generally speaking, the Rashba-effect at interfaces, even without inserted heavy metals, is expected to be present simply because structural inversion symmetry is broken \cite{linder_prl_11}. This makes the description of spin-orbit coupled interfaces of general interest in any type of hybrid structures comprised of two or more materials.

When it comes to describing quantum phenomena emerging at superconducting interfaces, the quasiclassical theory of superconductivity \cite{serene_physrep_83, belzig_sm_99, rammer_rmp_86} has proven useful and to a large degree consistent with experimental results. This theory is capable of treating non-equilibrium phenomena both in the ballistic and diffusive limit by obtaining the quasiclassical Green function $\check{g}$ in Keldysh-space. To achieve this aim, it is necessary to supplement the theory with boundary conditions at interfaces. For the case of interfaces with significant spin-orbit coupling, we have recently derived such boundary conditions in the diffusive limit~\cite{amundsen_prb_19}.  
These boundary conditions for $\check{g}$ correctly predict the appearance of triplet superconductivity in superconductors with spin-orbit coupled interfaces, but fail to capture the effect of motion parallel to the interface. While these contributions often cancel in the diffusive limit due to the momentum averaging brought on by frequent impurity scattering, an important special case when it does not is when a supercurrent is flowing parallel to the interface. In this case, gradient terms in the boundary conditions are required in order to capture the possible existence of spin-Hall like phenomena involving supercurrent flow and spin accumulation. Whereas such effects are of high interest experimentally, an accompanying theoretical description requires a derivation of boundary conditions that include such gradient terms. 

Here, we derive a boundary condition for spin-orbit coupled interfaces in the quasiclassical theory of superconductivity that includes precisely such gradient terms. This enables the description of spin-Hall like effects with superconductors due to interfaces with \eg Rashba-like spin-orbit coupling. After deriving these boundary conditions, we apply them to predict that a supercurrent flowing through a superconductor that is coupled to a normal metal via a spin-orbit interface can induce a non-local magnetization in the normal metal. This constitutes a superconducting equivalent of the conventional, resistive spin-Hall effect where a charge current induces a transverse spin accumulation. The boundary conditions derived here enables the exploration of quantum phenomena that emerge in superconductors due to spin-orbit coupled interfaces, both in and out of equilibrium.

\section{Derivation of boundary conditions}

We split the derivation of the boundary conditions into two parts: one part that takes into account the effect of tunneling through the interface and one part that takes into account reflection at the interface.

\subsection{Terms related to tunneling through the interface}

We consider a Hamiltonian of the form
\begin{align}
H = H_L + H_R + H_T,
\end{align}
with
\begin{align}\label{eq:HT}
H_{L} =& \sum_{\pv\pv'ss'} H^{L}_{ss'}(\pv,\pv')a^\dagger_{\pv s}a_{\pv's'}, \\
H_{R} =& \sum_{\pv\pv'ss'} H^{R}_{ss'}(\pv,\pv')b^\dagger_{\pv s}b_{\pv's'}, \\
H_T   =&  \sum_{\pv\pv'ss'} [T_{ss'}(\pv,\pv') a^\dagger_{\pv s} b_{\pv's'} + T^*_{ss'}(\pv,\pv')b^\dagger_{\pv's'}a_{\pv s}].
\end{align}
From Heisenbergs equation, one obtains for the operators
\begin{align}
i\hbar\partial_t a_{\pv s} =& +\sum_{\pv's'} [H^L_{ss'}(\pv,\pv')a_{\pv's'} + T_{ss'}(\pv,\pv')b_{\pv's'}], \notag\\
i\hbar\partial_t a^\dagger_{-\pv s} =& -\sum_{\pv's'} [[H^L_{ss'}(-\pv,-\pv')]^*a^\dagger_{-\pv's'} + T^*_{ss'}(-\pv,-\pv')b^\dagger_{-\pv's'}], \notag\\
i\hbar\partial_t b_{\pv s} =& +\sum_{\pv's'}[ H^R_{ss'}(\pv,\pv')b_{\pv's'}+ T^*_{s's}(\pv',\pv)a_{\pv's'}], \notag\\
i\hbar\partial_t b^\dagger_{-\pv s} =& -\sum_{\pv's'} [[H^R_{ss'}(-\pv,-\pv')]^*b^\dagger_{-\pv's'} + T_{s's}(-\pv',-\pv)b^\dagger_{-\pv's'} ].
\end{align}
Using Nambu vectors, $A_{\pv} = \begin{pmatrix} a_{\pv\uparrow} & a_{\pv\downarrow} & a^\dagger_{-\pv\uparrow} & a^\dagger_{-\pv\downarrow}\end{pmatrix}$ and similarly for $B_{\pv}$, this can be written as
\begin{align}
i\hbar\hat{\rho}_3\partial_t A_{\pv} =& \sum_{\pv'} [ \hat{H}^L(\pv,\pv') A_{\pv'} + \hat{T}(\pv,\pv')B_{\pv'} ],\\
i\hbar\hat{\rho}_3\partial_t B_{\pv} =& \sum_{\pv'} [ \hat{H}^R(\pv,\pv') B_{\pv'} + \hat{\tilde{T}}(\pv,\pv')A_{\pv'} ],
\end{align}
where $\hat{\rho}_3 = \text{diag}(1,1,-1,-1)$ and

\begin{align}
\hat{H}^{L/R}(\pv,\pv') =& \begin{pmatrix} H^{L/R}(\pv,\pv') & 0 \\ 0 & {H^{L/R}}^*(-\pv,-\pv') \end{pmatrix},\label{eq:Hn} \\
\hat{T}(\pv,\pv') =& \begin{pmatrix} T(\pv,\pv') & 0 \\ 0 & T^*(-\pv,-\pv')\end{pmatrix}, \\
\hat{\tilde{T}}(\pv,\pv') &= \begin{pmatrix} \tilde{T}(\pv,\pv') & 0 \\ 0 & \tilde{T}^*(-\pv,-\pv')\end{pmatrix}, 
\end{align}
where $T$ and $\tilde{T}$ are $2\times2$ matrices in spin space and we have defined the elements $\tilde{T}_{ss'}(\pv,\pv') \equiv T^*_{s's}(\pv',\pv)$. The tunneling current from the left to the right side is given as $J = e \langle \partial_t N \rangle$, with $N = \sum_{\pv s} a^\dagger_{\pv s}a_{\pv s}$. From the Heisenberg equation we find
\begin{align}
i\hbar\partial_t N &=  [ N,\; H ] =  [ N,\; H_T ] \notag\\
&= \sum_{\pv\pv'ss'} \Big[T_{ss'}(\pv,\pv')a^\dagger_{\pv s}b_{\pv's'} - T^*_{ss'}(\pv,\pv') b^\dagger_{\pv's'}a_{\pv s}\Big].
\end{align}
We note that while we have here written $H_{L,R}$ without superconducting terms, for notational simplicity, the approach generalizes straightforwardly to superconducting systems by adding off-diagonal terms to \cref{eq:Hn}. The following derivation of the boundary conditions for the Green function is valid irrespective of whether superconducting terms are included in $\hat{H}^{L/R}$ or not. Next, we define the Green function
\begin{align}
C^<_{ss'}(\pv,t;\pv',t') = i \langle a^\dagger_{\pv's'}(t') b_{\pv s}(t) \rangle,
\end{align}
which gives
\begin{align}\label{eq:n0}
\langle \partial_t N\rangle = -\frac{1}{\hbar}\sum_{\pv\pv'ss'} &\Big[T_{ss'}(\pv,\pv')C^<_{s's}(\pv',t;\pv,t) \notag\\
&+ T^*_{ss'}(-\pv,-\pv')i \langle a_{-\pv s} b^\dagger_{-\pv 's'} \rangle\Big].
\end{align}
Notice now that if we define a Nambu Green function matrix as
\begin{align}
\hat{C}^<(\pv,t;\pv',t') = i \langle [ [A^\dagger_{\pv'}(t') ]^T [B_{\pv}(t) ]^T ]^T \rangle,
\end{align}
the angular brackets in the second term of \cref{eq:n0} become the hole part of $\hat{C}^<$. Hence, we may write
\begin{align}
 \langle\partial_t N \rangle = -\frac{1}{\hbar}\sum_{\pv\pv'} \Tr [\hat{T}(\pv,\pv')\hat{C}^<(\pv',t;\pv,t) ].
\label{eq:dtn}
\end{align} 
For later use, we also introduce the Green function $\hat{C}^>$ and Keldysh Green function $\hat{C}^K$ as:
\begin{align}
\hat{C}^>(\pv,t;\pv',t') &=-i\left\langle B_{\pv}(t)A^\dagger_{\pv'}(t')\right\rangle,\notag\\
\hat{C}^K(\pv,t;\pv',t') &= -i \hat{\rho}_3 \langle [B_{\pv}(t), A^\dagger_{\pv'}(t')]\rangle.
\end{align}
To find $\hat{C}^<$, we first find its time-ordered equivalent,
\begin{align}
\hat{C}(\pv,t;\pv',t') =& -i\hat{\rho}_3 \langle\Tm    B_{\pv}(t) A^\dagger_{\pv'}(t') \rangle \nonumber\\ 
=& -i\hat{\rho}_3\theta(t-t') \langle  B_{\pv}(t)A^\dagger_{\pv'}(t') \rangle \notag\\
&+i\hat{\rho}_3\theta(t'-t) \langle  [[A^\dagger_{\pv'}(t')]^T[B_{\pv}(t)]^T]^T  \rangle
\end{align}
where $\Tm$ is the time-ordering operator, which acts as a normal-operator at equal times. Differentiating with respect to $t$ gives us
\begin{align}
i\hbar\hat{\rho}_3\partial_t\hat{C}(\pv,t;\pv',t') &= \sum_{\qv} \Big[\hat{H}^R(\pv,\qv)\hat{C}(\qv,t;\pv',t') \notag\\
&+ \hat{\tilde{T}}(\pv,\qv)\hat{G}_L(\qv,t;\pv',t')\Big],
\end{align}
where  
\begin{align*}
\hat{G}_L(\qv,t;\pv',t') = -i\hat{\rho}_3 \langle\Tm A_{\qv}(t) A^\dagger_{\pv'}(t')  \rangle.
\end{align*}
This may be written as
\begin{align}
\sum_{\qv} (i\hbar\hat{\rho}_3\partial_t\delta_{\qv\pv} - &\hat{H}^R(\pv,\qv) )\hat{C}(\qv,t;\pv',t')\notag\\
&= \sum_{\qv}\hat{\tilde{T}}(\pv,\qv)\hat{G}_L(\qv,t;\qv',t').
\end{align}
A solution to this equation is given by
\begin{align}
\hat{C}(\pv,t;\pv',t') &= \int dt_1\sum_{\qv\qv'}\hat{G}_{0,R}(\pv,t;\qv,t_1)\hat{\tilde{T}}(\qv,\qv') \notag\\
&\times\hat{G}_L(\qv',t_1;\pv',t'),
\end{align}
with $\hat{G}_{0,R}$ the right-side unperturbed Green function, satisfying
\begin{align}
\sum_{\qv} (i\hbar\hat{\rho}_3\partial_t\delta_{\pv\qv} - \hat{H}^R(\pv,\qv) )\hat{G}_{0,R}(\qv,t;\pv',t') = \delta(t-t')\delta_{\pv\pv'},
\end{align}
as can be seen by direct insertion. Next, we deform the time axis to the Keldysh contour,

\begin{align}
\hat{C}(\pv,\tau;\pv',\tau') &= \int_C d\tau_1\sum_{\qv\qv'}\hat{G}_{0,R}(\pv,\tau;\qv,\tau_1)\hat{\tilde{T}}(\qv,\qv') \notag\\
&\times \hat{G}_L(\qv',\tau_1;\pv',\tau'),
\end{align}

to obtain via the Langreth rules \cite{jauho}

\begin{widetext}
\begin{align}\label{eq:clesgreat}
\hat{\rho}_3\hat{C}^<(\pv,t;\pv',t') =& \int dt_1\sum_{\qv\qv'} [\hat{G}^R_{0,R}(\pv,t;\qv,t_1)\hat{\tilde{T}}(\qv,\qv')\hat{\rho}_3 \hat{G}^<_L(\qv',t_1;\pv',t') + \hat{\rho}_3\hat{G}^<_{0,R}(\pv,t;\qv,t_1)\hat{\tilde{T}}(\qv,\qv')\hat{G}^A_L(\qv',t_1;\pv',t') ], \\
\hat{\rho}_3\hat{C}^>(\pv,t;\pv',t') =& \int dt_1\sum_{\qv\qv'} [\hat{G}^R_{0,R}(\pv,t;\qv,t_1)\hat{\tilde{T}}(\qv,\qv')\hat{\rho}_3\hat{G}^>_L(\qv',t_1;\pv',t') + \hat{\rho}_3\hat{G}^>_{0,R}(\pv,t;\qv,t_1)\hat{\tilde{T}}(\qv,\qv')\hat{G}^A_L(\qv',t_1;\pv',t') ].
\end{align}
Note that the lesser and greater Green functions were defined without $\hat{\rho}_3$, whereas the retarded, advanced and time-ordered were defined with. Hence, when going from time-ordered to lesser/greater, there has to be additional $\hat{\rho}_3$ appearing, as shown in the above equations. The lesser and greater Green functions are further related to the Keldysh Green function as
\begin{align}
\hat{G}^K(\pv,t;\pv',t') = \hat{\rho}_3 [\hat{G}^<(\pv,t;\pv',t') + \hat{G}^>(\pv,t;\pv',t')].
\end{align}
The same relation also holds if one replaces $\hat{G}$ with $\hat{C}$. We also have the relationship
\begin{align}
\hat{C}^<(\pv,t;\pv',t) = \hat{C}^>(\pv,t;\pv',t),
\end{align}
By adding Eqs. (\ref{eq:clesgreat}) and then multiplying the whole equation with $\hat{\rho}_3/2$, one therefore obtains:
\begin{align}
\hat{C}^<(\pv,t;\pv',t) =& \frac{1}{2}\hat{\rho}_3\int dt_1\sum_{\qv\qv'} [\hat{G}^R_{0,R}(\pv,t;\qv,t_1)\hat{\tilde{T}}(\qv,\qv')\hat{G}^K_L(\qv',t_1;\pv',t) + \hat{G}^K_{0,R}(\pv,t;\qv,t_1)\hat{\tilde{T}}(\qv,\qv')\hat{G}^A_L(\qv',t_1;\pv',t) ].
\end{align}
\end{widetext}
Insertion into \cref{eq:dtn} then gives
\begin{align}
 \langle \partial_t N  \rangle &= -\frac{1}{2\hbar}\sum_{\pv\pv'\qv\qv'}\int dt_1\Tr [\hat{\rho}_3 (\hat{T}(\pv,\pv')\check{G}_{0,R}(\pv',t;\qv,t_1) \notag\\
&\times\hat{\tilde{T}}(\qv,\qv')\check{G}_L(\qv',t_1;\pv,t) )^K ],
\end{align}
where it has been used that $\hat{T}$ commutes with $\hat{\rho}_3$. The superscript $\emph{K}$ means in this context that the Keldysh component of the matrix product enclosed in the parentheses should be extracted. Repeating the process by calculating the currents flowing into the opposite material gives
\begin{align}
 \langle \partial_t N  \rangle &= +\frac{1}{2\hbar}\sum_{\pv\pv'\qv\qv'}\int dt_1\Tr\Big[\hat{\rho}_3 (\check{G}_{0,L}(\pv',\qv)\hat{T}(\qv,\qv')\notag\\
&\times \check{G}_{R}(\qv',t;\pv,t_1)\hat{\tilde{T}}(\pv,\pv') )^K\Big].
\end{align}
Averaging the two and Fourier transforming then gives
\begin{align}
 \langle \partial_t N  \rangle =&+\frac{1}{4\hbar}\sum_{\pv\pv'\qv\qv'}\int \frac{d\varepsilon}{2\pi}\Tr\Big[\hat{\rho}_3 (\check{G}_{0,L}(\pv,\qv;\varepsilon)\hat{T}(\qv,\qv') \notag\\
&\times \check{G}_R(\qv',\pv';\varepsilon)\hat{\tilde{T}}(\pv',\pv) )^K\Big] \nonumber\\
&-\frac{1}{4\hbar}\sum_{\pv\pv'\qv\qv'}\int \frac{d\varepsilon}{2\pi}\Tr\Big[\hat{\rho}_3 (\hat{T}(\pv',\pv)\check{G}_{0,R}(\pv,\qv;\varepsilon) \notag\\
&\times \check{\hat{T}}(\qv,\qv')\check{G}_L(\qv',\pv';\varepsilon) )^K\Big]
\label{eq:bcbase}
\end{align}

where we defined the $8\times 8$ matrix $\check{G}$ in Keldysh-space:
\begin{align}
\check{G} = \begin{pmatrix}
\hat{G}^R & \hat{G}^K \\
\hat{0} & \hat{G}^A \\
\end{pmatrix}.
\end{align}
A product $\hat{A}\check{B}$ between a $\hat{A}$-matrix and $\check{B}$-matrix is to be understood as $\check{A}\check{B}$ where $\check{A} = \text{diag}(\hat{A},\hat{A})$.
In quasiclassical theory, the electric current $J$ is computed from a quantity known as the matrix current $\check{I}$ via the expression:
\begin{align}
J \propto \int dE\; \text{Tr}\{\hat{\rho}_3 \check{I}^K\}
\end{align}
Therefore, one may identify the matrix current as
\begin{align}
\check{I} \propto \sum_{\pv\pv'\qv\qv'} [\check{G}_L, \hat{T}\check{G}_R\hat{\tilde{T}}] + \text{traceless terms}.
\end{align}
where we has set approximated $\check{G}_{0,j}\simeq \check{G}_j$ for material $j$. The designation "traceless terms" indicates that there may exist terms in the matrix current which do not give any contribution to the charge current, in effect which give zero when multiplied with $\hat{\rho}_3$ and taken the trace over. In the next subsection, we shall identify these terms.

To compute the boundary conditions specific for an interface with Rashba spin-orbit coupling, we assume an interface potential of the form
\begin{align}
T(\rv) =  T_0(\rv_\para)\delta(r_\perp - R_0) -i\{T_{ij}(\rv_\para)\delta(r_\perp - R_0)\sigma_j,\partial_i\} ,
\end{align}
where $r_\para$ and $r_\perp$ are coordinates parallel and perpendicular to the interface. Note that the term involving the differential operator $\partial_i$ has been symmetrized in order to ensure hermiticity. The tunneling Hamiltonian is 

\begin{align}
H_T = \sum_{ss'}\int d\rv\;  a_s^\dagger(\rv)T_{ss'}(\rv)b_{s'}(\rv) +  \text{h.c.}
\end{align}

with $a(\rv)$ and $b(\rv)$ the field operators on the left and right side of the interface, respectively. We may now define $a_s(\rv) = \sum_{\pv} a_{\pv s}e^{i\pv\cdot\rv}$, and $b_s(\rv) = \sum_{\pv} b_{\pv s}e^{i\pv\cdot\rv}$ which produces precisely $H_T$ in Eq. (\ref{eq:HT}) with
%
%

\begin{align}
T_{ss'}(\pv,\pv') &= \int d\rv_\para [T_0(\rv_\para) + T_{ij}(\rv_\para)(p_i+p_i')\sigma_{ss',j} ]\notag\\
&e^{-i(\pv-\pv')\cdot\rv_\para}.
\end{align}

We now approximate the tunneling matrix element as follows to describe the effect of interfacial Rashba-coupling:
\begin{align}
T_{ss'}(\pv,\pv')\simeq T_0 + T_{ij} (p_i+p_i')\sigma_{ss',j} ,
\label{eq:tmat}
\end{align}
where $T_0$ and $T_{ij}$ are real phenomenological constants. As an aside, this approximation is equivalent to saying that both $T_0$ and $T_{ij}$ have a $\delta(\rv_\para)$ spatial dependency, which means that the surface is approximated as a point impurity. Moreover, we note that from the definition of $\tilde{T}(\pv,\pv')$, it then follows that 
\begin{align}
\tilde{T}_{ss'}(\pv,\pv')\simeq T_0 + T_{ij}(p_i'+p_i)\sigma_{ss',j} = T_{ss'}(\pv,\pv'),
\end{align}

Next, we need to calculate the integral
\begin{align}
\check{M} &=\int d\vec{p} d\pv'd\qv d\qv'\;  \hat{T}(\pv',\pv)\check{G}_R(\vec{p},\qv)\hat{\tilde{T}}(\qv,\qv')\check{G}_L(\qv',\pv')
\end{align}
which is one of the two main terms appearing in Eq. (\ref{eq:bcbase}) after going to the continuum limit in momentum-space. Defining
\begin{align}
\hat{T}(\pv,\pv') &= T_0 + (p_i + p_i')\hat{t}_i,\notag\\
\hat{t}_i &= T_{ij}\hat{\rho}_3 \hat{\sigma}_j,\notag\\
\hat{\boldsymbol{\sigma}} &= \begin{pmatrix}
\boldsymbol{\sigma} & 0 \\
0 & \boldsymbol{\sigma}^* \\
\end{pmatrix}.
\end{align}
we get
\begin{align}
M &= \int d\vec{p} d\pv'd\qv d\qv'\; [T_0^2\check{G}_R(\pv,\qv)\check{G}_L(\qv',\pv') \notag\\
&+ T_0 \check{G}_R(\pv,\qv) \hat{t}_k (q_k + q_k') \check{G}_L(\qv',\pv') \notag\\
&+T_0\hat{t}_i \check{G}_R(\pv,\qv) (p_i+p_i') \check{G}_L(\qv',\pv') \notag\\
&+\hat{t}_i \check{G}_R(\pv,\qv) \hat{t}_k (p_i+p_i')(q_k+q_k') \check{G}_L(\qv',\pv')].
\end{align}

We make use of the fact that the Green function $\check{G}_{L/R}(\pv,\qv)$ is strongly peaked for $\pv \simeq \qv \simeq \vecp_F$, since the aim is to construct a low-energy theory relative the Fermi level. For such low energies, particle have momenta close to $\vecp_F$. Since both $\pv$ and $\qv$ are close to $\vecp_F$, the center-of-mass momentum $(\vecp+\vecq)/2$ must also be close to $\vecp_F$. Therefore, we may approximate the individual momenta variables $\{p_i,p_i',q_k,q_k'\}$ entering the above equation as center-of-mass variables with magnitude equal to the Fermi momentum. Making use of the fact that the Jacobian of a transformation from the momenta of the field operators to a center-of-mass and relative momentum representation is unity, we find within the quasiclassical approximation that
\begin{align}\label{eq:mpropto}
\check{M} &\propto T_0^2 \check{g}_{R,s}\check{g}_{L,s} - m_RD_RT_0 \check{g}_{R,s} \partial_k \check{g}_{R,s} \hat{t}_k \check{g}_{L,s} \notag\\
&-m_LD_LT_0 \check{g}_{R,s} \hat{t}_k \check{g}_{L,s} \partial_k \check{g}_{L,s} - m_RD_RT_0 \hat{t_i} \check{g}_{R,s}\partial_i \check{g}_{R,s} \check{g}_{L,s} \notag\\
&-m_LD_LT_0\hat{t}_i \check{g}_{R,s}\check{g}_{L,s} \partial_i \check{g}_{L,s} + \frac{2p_F^2}{3}\hat{t}_i \check{g}_{R,s}\hat{t}_i \check{g}_{L,s}.
\end{align}
To obtain this result, we used that
\begin{align}
\check{g}_{R/L}(\pv_F,\vecR=0) = \frac{1}{(2\pi)^3}\frac{i}{\pi} \int d\qv \int d\xi_\vecp \check{G}_{R/L}(\pv,\qv)
\end{align}
is the quasiclassical Green function evaluated at the interface and $\qv$ in the equation above is the relative momentum-coordinate whereas $\pv$ is the center-of-mass momentum. We also inserted $\check{g}_R = \check{g}_{R,s} + \hat{\boldsymbol{p}}_F\cdot \check{\bm{g}}_{R}$, where $\check{\bm{g}}_R = -\tau v_F \check{g}_{R,s}\nabla \check{g}_{R,s}$, as follows from the Usadel equation, and $\vecp$ is now the center-of-mass momentum coordinate. For the angular averaging, we used that
\begin{align*}
\int \frac{d\Omega_p}{4\pi} \; p_{F,i}p_{F,j}/p_F^2 = \frac{1}{3}\delta_{ij},
\end{align*}.
Here, $m$ is the electron mass and we used that $p_F = mv_F$ and $D = v_F^2\tau/3$. Since the parameters $T_0$ and $T_{ij}$ will be treated as phenomenological interface parameters, we absorb the proportionality constant in Eq. (\ref{eq:mpropto}) into them. We also absorb a factor 1/3 into $p_F^2$ in the last term of Eq. (\ref{eq:mpropto}) for brevity of notation. Finally, we neglected terms in Eq. (\ref{eq:mpropto}) which were second order in gradient terms of the Green functions under the assumption that such terms are small.

The boundary conditions thus become
\begin{align}
\boldsymbol{n}\cdot D\cg_L\nabla \cg_L &= T_0^2  [\cg_L,\; \cg_R ] + 2T_{ij}T_{il} p_F^2 [\cg_L\;,\; \hat{\rho}_3\hat{\sigma}_j \cg_R\hat{\rho}_3\hat{\sigma}_l ] \notag\\
&- mDT_{ij}T_0 [\cg_L,\;  \{\hat{\rho}_3\hat{\sigma}_j\;,\;\cg_R\partial_i \cg_R \} ] \notag\\
&- mDT_{ij}T_0 [\cg_L \partial_i \cg_L,\;  \{\hat{\rho}_3\hat{\sigma}_j\;,\;\cg_R \} ]
\end{align}
For the special case of Rashba SOC, we have $T_{ij} = T_1n_k\varepsilon_{kji}$, where $n_k$ is component $k$ of the interface normal. This gives
\begin{align}\label{eq:rashbagradient}
\boldsymbol{n}\cdot D\cg_L\nabla \cg_L &= T_0^2  [\cg_L,\; \cg_R ] + 2T_1^2 p_F^2[\cg_L,\; \bm{\sigma}_{||} \cg_R\bm{\sigma}_{||} ]\notag\\
 &- mDT_1T_0 [\cg_L,\;  \{\bm{\sigma}_{||},\;(\cg_R\nabla \cg_R)\times\boldsymbol{n} \} ] \notag\\
&- mDT_1T_0 [(\cg_L\nabla \cg_L)\times\boldsymbol{n},\;  \{\bm{\sigma}_{||},\; \cg_R\} ].
\end{align}
with $\hat{\bm{\sigma}}_{||} = \hat{\rho}_3 [\hat{\bm{\sigma}} - \boldsymbol{n}(\hat{\bm{\sigma}}\cdot\boldsymbol{n}) ]$. We see that the second term on the right-hand side of Eq. (\ref{eq:rashbagradient}) reproduces the tunneling term due to spin-orbit coupling derived in Ref. \cite{amundsen_prb_19}. The third and fourth terms on the right-hand side is new and involves the gradient of the Green function. This allows us to describe for instance spin-Hall like phenomena with supercurrents, as we will demonstrate below.

\subsection{Terms related to reflection at the interface}

\begin{figure}[b!]
  \centering
  \includegraphics[width=1.0\linewidth]{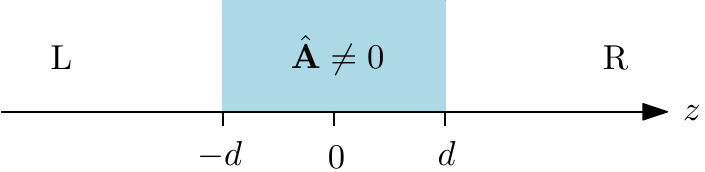}
  \caption{(Color online) Spin-orbit coupled interface separating left (L) and right (R) region in the model used to derive the reflection boundary condition terms. 
  }%
  \label{fig:reflectionsetup}
\end{figure}

Here, we follow a similar procedure as \cite{bergeret_prb_12, silaev_prb_20} to derive the terms related to reflection at the interface. The Usadel equation including antisymmetric spin-orbit coupling reads:
\begin{align}
D \tnabla (\cg \tnabla \cg) + i[\epsilon\hat{\rho}_3 + \hat{\Sigma}, \cg] = 0.
\end{align}
Here, we have defined
\begin{align}
\tnabla \cdot \ldots = \nabla \cdot \ldots - i [\bha, \ldots].
\end{align}
We use the convention that when multiplying a matrix $\hat{A}$ ($4 \times 4$) with a matrix $\check{B}$ ($8\times 8$), that product should be understood as $\text{diag}(\hat{A}, \hat{A}) \cdot \check{B}$. The matrix $\bha$ has vectors as elements and classifies the spin-orbit coupling according to
\begin{align}
\bha = \begin{pmatrix}
\vecA & 0 \\
0 & -\vecA^* \\
\end{pmatrix}
\end{align}
where $\vecA$ defines the spin-orbit coupling part of the Hamilton-operator via:
\begin{align}
H_\text{soc} = -\frac{1}{m} \veck \cdot \vecA.
\end{align}
This is a generally valid Hamilton-operator for an antisymmetric spin-orbit coupling that is linear in momentum. For instance, the Rashba case $H_\text{soc} = -\frac{\alpha}{m} (\veck\times\vecsigma)\cdot \vecz$ for structures breaking inversion symmetry in the $z$-direction is obtained using $\vecA = \alpha(-\sigma_y, \sigma_x,0)$. Here, we will derive boundary conditions valid for any $\vecA$ that satisfies $\vecn\cdot\vecA=0$ where $\vecn$ is the interface normal vector. This includes the Rashba interaction as a limiting case. The criterion $\vecn\cdot\vecA=0$ is chosen as to avoid the complication of symmetrizing the spin-orbit coupling Hamiltonian, as is required for the hermiticity of the Hamilton-operator when the spin-orbit coupling is confined to a certain spatial region. Nevertheless, the condition $\vecn\cdot\vecA$ still allows for several interesting new phenomena related to spin-Hall physics to be predicted.\\

The underlying assumption is that the region with spin-orbit coupling is sufficiently thin that we may neglect the spatial variation of the Green function along the interface normal $\vecn$. Let $n$ denote the coordinate along the $\vecn$-axis (chosen as $z$-direction in Fig. \ref{fig:reflectionsetup}). Integrating the Usadel-equation from $n=-d$ to $n=d$ gives:
\begin{align}\label{eq:currenttemp1}
(\cg\partial_n& \cg)|_d - (\cg\partial_n\cg)|_{-d} - i d\nabla \cg[\hat{\vecA}, \cg]|_d \notag\\
&- i \check{g} d [\hat{\vecA},\nabla{\check{g}}]- i d[\hat{\vecA},\cg\nabla\cg]|_d - d[\hat{\vecA}, \cg [\hat{\vecA},\cg]]|_d = 0.
\end{align}
Here, we made use of the fact that $\vecn\cdot\vecA=0$ and approximated the value of all terms not involving a derivative in the $n$-direction with their value at $n=d$. This is consistent with the assumption that $d$ is smaller than the length scale over which the Green function varies in the $n$-direction. Now, if there was no spin-orbit coupling present, the tunneling current alone derived in the previous section (with scalar tunneling amplitudes due to the absence of SOC) provides the value of $\cg \partial_n \cg|_d$, constituting the effect boundary condition for the Green function $\cg$. Therefore, we may identify:
\begin{align}
(\cg\partial_n\cg)|_{-d} = \text{tunneling current contribution} \equiv I_T,
\end{align}
and consequently write Eq. (\ref{eq:currenttemp1}) as:
\begin{align}
\cg \partial_n& \cg|_d = I_T + i d \nabla\cg[\hat{\vecA},\cg]_d + i d\cg [\hat{\vecA},\nabla\cg]_d \notag\\
&+ i d[\hat{\vecA}, \cg\nabla\cg]_d + d[\hat{\vecA}, \cg[\hat{\vecA},\cg]]_d.
\end{align}
After cancelling some terms, we end up with the final form which is generally valid for any SOC at the interface so long that $\vecn\cdot\vecA=0$:
\begin{align}\label{eq:currenttemp2}
\boldsymbol{n}\cdot \cg \nabla& \cg|_d = I_T + i d\nabla\cg[\hat{\vecA}, \cg]|_d - 2i d\cg\nabla\cg \hat{\vecA}_d + i d\cg\hat{\vecA}\nabla\cg|_d \notag\\
&+ i d\hat{\vecA}\cg\nabla\cg_d + d [\hat{\vecA}, \cg\hat{\vecA}\cg]_d.
\end{align}
This constitutes the boundary condition for the Green function on the right side of the interface $\check{g}_R = \check{g}|_d$. The boundary condition for the Green function on the left side of the interface $\check{g}_L = \check{g}|_{-d}$ is obtained from Eq. (\ref{eq:currenttemp2}) by letting $d \to -d$. This is derived in exactly the same way we did for $\check{g}_R$ except that we integrate $\int_d^{-d}$ instead of $\int_{-d}^d$. \\

\begin{figure}[b!]
  \centering
  \includegraphics[width=1.0\linewidth]{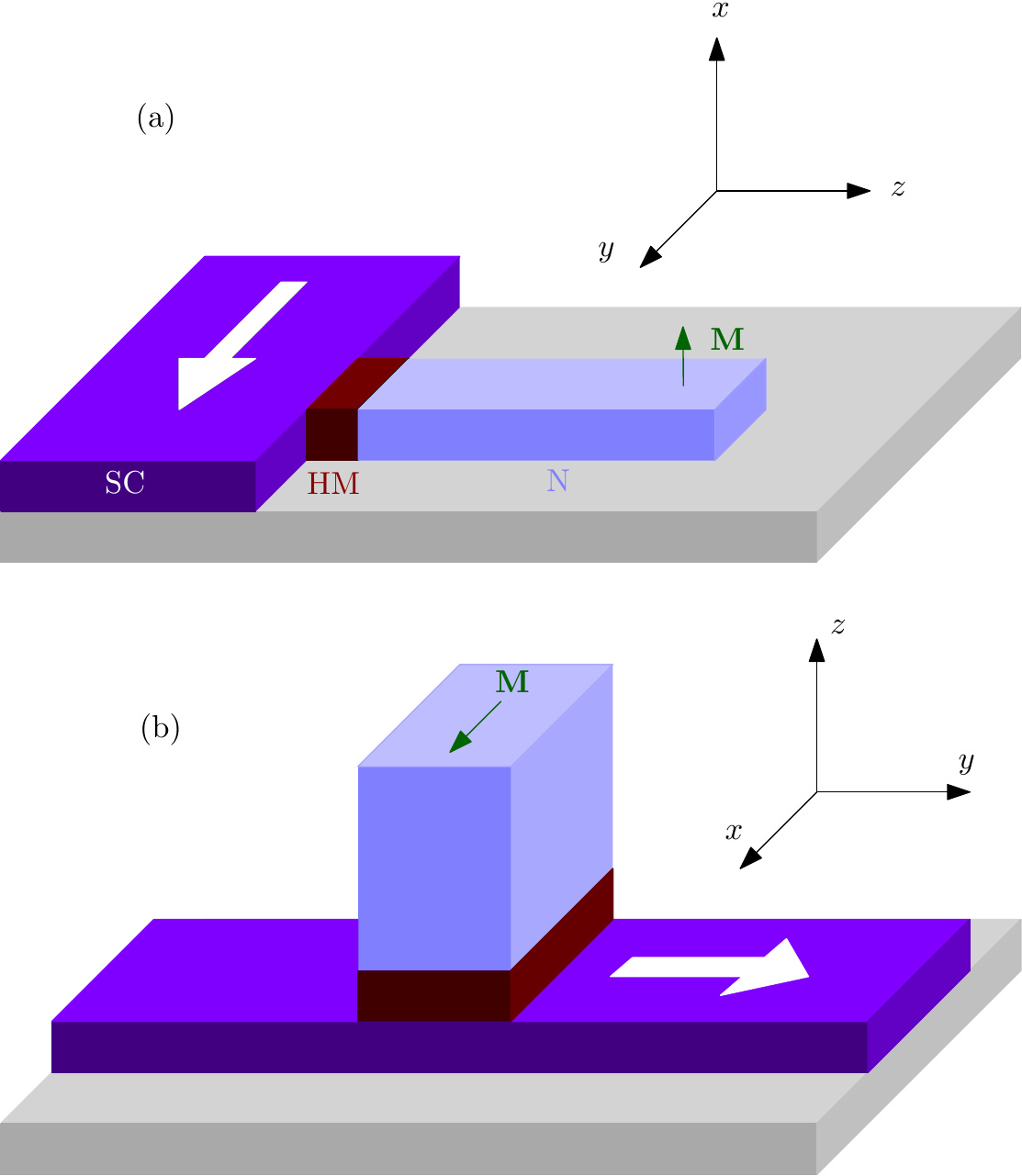}
  \caption{(Color online) Two possible experimental setups for observation
of supercurrent-induced magnetization via a Rashba interface. In
both cases, current-biasing the superconductor induces a non-local
spin magnetization in a normal metal without any magnetic elements
in the system. The magnetization points in the direction orthogonal to both the supercurrent flow and the broken inversion symmetry axis, similarly to the conventional spin Hall effect.
  }%
  \label{fig:socsetup}
\end{figure}

As an application of these boundary conditions, consider now a special case. We examine a 2D structure ($yz$-plane) with a Rashba spin-orbit coupled interface. The $z$-direction is perpendicular to the interface plane, so that $\boldsymbol{n} = \hat{\boldsymbol{z}}$. We permit not only $\partial_z \cg_R \neq 0$, but also $\partial_y \cg_R \neq 0$ in order to capture spin Hall like phenomena involving motion parallell with the interface. In the Rashba case, we have as previously mentioned $\vecA = \alpha(-\sigma_y,\sigma_x,0)$, which gives
\begin{align}
\cg_R \partial_z& \cg_R = \check{I}_T + i d\partial_y \cg_R[\hat{A}_y,\cg_R] - 2i d \cg_R \partial_y \cg_R \hat{A}_y + i d \cg_R \hat{A}_y \partial_y \cg_R\notag\\
&+ i d \hat{A}_y \cg_R \partial_y \cg_R + d[\hat{\vecA}, \cg_R \hat{\vecA} \cg_R].
\end{align}
We define
\begin{align}
\hat{\rho}_x = \begin{pmatrix}
\sigma_x &  0 \\
0 & -\sigma_x^* \\
\end{pmatrix},\; 
\hat{\rho}_y = \begin{pmatrix}
\sigma_y & 0 \\
0 & - \sigma_y^* \\
\end{pmatrix},
\end{align}
so that $\hat{A}_x = -\alpha\hat{\rho}_y$ and $\hat{A}_y = \alpha\hat{\rho}_x$. We then get:
\begin{widetext}
\begin{align}
\cg_R \partial_z \cg_R &= \check{I}_T + i d\alpha\partial_y \cg_R[\hat{\rho}_x,\cg_R] -2i d\alpha \cg_R\partial_y \cg_R\hat{\rho}_x + i d\alpha \cg_R\hat{\rho}_x\partial_y \cg_R +i d\alpha \hat{\rho}_x \cg_R \partial_y \cg_R \notag\\
&+ d\alpha^2[\hat{\rho}_x,\cg_R\hat{\rho}_x \cg_R] + d\alpha^2[\hat{\rho}_y,\cg_R\hat{\rho}_y\cg_R].
\end{align}
It is clear from the boundary conditions that any variation in the Green function $\check{g}_R$ in the direction parallell to the interface ($y$-direction) will now couple directly to variations of the Green function in direction normal to the interface ($z$-direction) due to the Rashba spin-orbit coupling at the interface.

\subsection{Final form of complete boundary condition for Rashba-coupling}
In the preceeding sections, we have derived the contribution to the boundary condition for the quasiclassical Green function coming both from tunneling terms and reflection terms at an interface with spin-orbit coupling. Here, we give the final complete form of the boundary conditions for the Green function on both the left and right side of the interface. Let $\boldsymbol{n}$ point from left to right.

For the quasiclassical Green function matrix $\check{g}_L$ on the left side of the interface, the boundary condition generally reads
\begin{align}
\boldsymbol{n}\cdot D\cg_L\nabla \cg_L &= T_0^2  [\cg_L,\; \cg_R ] - mDT_{ij}T_0 [\cg_L\;,\;  \{\hat{\rho}_3 \hat{\sigma}_j\;,\;\cg_R\partial_i \cg_R \} ] \;- mDT_{ij}T_0 [\cg_L \partial_i \cg_L\;,\;  \{\hat{\rho}_3 \hat{\sigma}_j\;,\;\cg_R \} ] + T_{ij}T_{il}p_F^2 [\cg_L\;,\; \hat{\rho}_3 \hat{\sigma}_j \cg_R\hat{\rho}_3 \hat{\sigma}_l ] \notag\\
&- i Dd\nabla\cg_L[\hat{\vecA}, \cg_L] + 2i Dd\cg_L\nabla\cg_L \hat{\vecA} - i Dd\cg_L\hat{\vecA}\nabla\cg_L - i Dd\hat{\vecA}\cg_L\nabla\cg_L - Dd [\hat{\vecA}, \cg_L\hat{\vecA}\cg_L].
\end{align}
while for the Green function on the right side:
\begin{align}
\boldsymbol{n}\cdot D\cg_R\nabla \cg_R &= T_0^2  [\cg_L,\; \cg_R ] + mDT_{ij}T_0 [\cg_R\;,\;  \{\hat{\rho}_3\hat{\sigma}_j\;,\;\cg_L\partial_i \cg_L \} ] \;+ mDT_{ij}T_0 [\cg_R \partial_i \cg_R\;,\;  \{\hat{\rho}_3 \hat{\sigma}_j\;,\;\cg_L \} ] - T_{ij}T_{il} p_F^2[\cg_R\;,\; \hat{\rho}_3\hat{\sigma}_j \cg_L\hat{\rho}_3\hat{\sigma}_l ] \notag\\
&+ i Dd\nabla\cg_R[\hat{\vecA}, \cg_R] - 2i Dd\cg_R\nabla\cg_R \hat{\vecA} + i Dd\cg_R\hat{\vecA}\nabla\cg_R + i Dd\hat{\vecA}\cg_R\nabla\cg_R + Dd [\hat{\vecA}, \cg_R\hat{\vecA}\cg_R].
\end{align}
For the special case of Rashba spin-orbit coupling and an interface normal $\boldsymbol{n}$ along the $z$-direction with $\check{g}_L = \check{g}_L(y)$ due to supercurrent flow in the $y$-direction and a thin region to the right of the interface so that $\check{g}_R = \check{g}_R(z)$, as shown in Fig. \ref{fig:socsetup}, one obtains on the left side:
\begin{align}
D\cg_L\partial_z \cg_L &= T_0^2  [\cg_L,\; \cg_R ] + T_1^2p_F^2 [\cg_L\;,\; \hat{\bm{\sigma}}_{||} \cg_R \hat{\bm{\sigma}}_{||} ]- mDT_1T_0 [\cg_L\;,\;  \{ \hat{\sigma}_{||,x}\;,\;\cg_R\partial_y \cg_R\} ] - mDT_1T_0 [\cg_L \partial_y\cg_L\;,\;  \{ \hat{\sigma}_{||,x}\;,\;\cg_R\} ]\notag\\
&- Dd\alpha^2[\hat{\rho}_x,\cg_L\hat{\rho}_x \cg_L] - Dd\alpha^2[\hat{\rho}_y,\cg_L\hat{\rho}_y\cg_L],
\end{align}
while on the right side:
\begin{align}
D\cg_R\partial_z \cg_R &= T_0^2  [\cg_L,\; \cg_R ] - T_1^2 p_F^2[\cg_R\;,\; \hat{\bm{\sigma}}_{||} g_L\hat{\bm{\sigma}}_{||} ]+ mDT_1T_0 [\cg_R\;,\;  \{\hat{\sigma}_{||,x}\;,\;\cg_L\partial_y \cg_L \} ] + mDT_1T_0 [\cg_R \partial_y\cg_R\;,\;  \{ \hat{\sigma}_{||,x}\;,\;\cg_L\} ]\notag\\
&+ Dd\alpha^2[\hat{\rho}_x,\cg_R\hat{\rho}_x \cg_R] + Dd\alpha^2[\hat{\rho}_y,\cg_R\hat{\rho}_y \cg_R].
\end{align}
In this particular case, one has $\boldsymbol{\hat{\sigma}}_{||} = (\hat{\rho}_x,\hat{\rho}_y,\hat{0})$. The derivation of the boundary conditions provided here treats the interface in a phenomenological manner, in particular when it comes to the reflection terms. A stricter microscopic derivation makes use of concepts like isotropization zone and ballistic zones near an interface \cite{cottet_prb_09}. Nevertheless, the methodology used in this paper to derive the boundary conditions is known to give the same results as the microscopic derivation in the case of magnetic boundary conditions \cite{bergeret_prb_12}, a fact which motivates the present approach.
\end{widetext}

\section{Application: supercurrent-induced magnetization via Rashba interface}

We here show how the boundary conditions can be used on
a concrete system and that they predict that a supercurrent can
induce a non-local magnetization via structural inversion symmetry
breaking. The polarization of the induced magnetization
is perpendicular to both the current flow and the direction of
inversion symmetry breaking and can be viewed as a type of
superconducting spin Hall effect.
We consider a superconductor in contact with a normal metal
through a thin heavy metal interface. Two possible experimental
realizations of this is shown in Fig. 1. When a supercurrent
runs through the superconductor, a non-local magnetization
appears in the normal metal despite the absence of any magnetic
elements in the system. To show this analytically, we consider
the weak proximity effect regime where the Usadel equation
is linearized in the anomalous Green function.

The quasiclassical retarded Green function in a supercurrent-carrying superconductor is given by
\begin{align}
\hat{g}_S = \begin{pmatrix}
c & s\e{i Jy}i \sigma_y\\
s\e{-i Jy}i \sigma_y & -c\\
\end{pmatrix}
\end{align}
for a current density $J$ well below the critical supercurrent density. Here, $c=\cosh(\theta)$, $s=\sinh(\theta)$, where $\theta=\text{atanh}(\Delta/\epsilon)$ where $\Delta$ is the superconducting gap. In the normal metal, the retarded Green function matrix reads
\begin{align}
\hat{g}_N = \begin{pmatrix}
1 & f\\
-\tilde{f} & -1\\
\end{pmatrix}
\end{align}
in the case of a weak proximity effect where the anomalous Green function matrix has the form
\begin{align}
f = \begin{pmatrix}
f_\uparrow & f_t + f_s \\
f_t - f_s & f_\downarrow \\
\end{pmatrix}.
\end{align}
The $\tilde{\ldots}$ operator denotes reversal of energy ($\varepsilon \to -\varepsilon$) and complex conjugation. In the normal metal, the Usadel equation then takes the form
\begin{align}
D\partial_z^2 f_{s,t,\sigma} + 2i\varepsilon f_{s,t,\sigma} = 0
\end{align}
and has a general solution
\begin{align}
f_{s,t,\sigma} = A_{s,t,\sigma} \e{i kz} + B_{s,t,\sigma} \e{-i kz}
\end{align}
where $k=\sqrt{2i\varepsilon/D}$. The boundary condition at the vacuum edge $z=L$ of the normal metal is
\begin{align}
\partial_z f_{s,t,\sigma} = 0
\end{align}
in order to ensure that no current flows into vacuum. The boundary condition between the normal metal and the superconductor at $z=0$, which are connected via an atomically thin Pt layer, is the key ingredient for achieving the superspin Hall effect. In the weak proximity effect regime, we obtain
\begin{align}
&D \partial_z f_\sigma + 4i\sigma DJT_0T_1csm \e{i Jy_0} - 2T_0^2 cf_\sigma \notag\\
&\;\;- 4i\sigma DmT_1Js^2T_0 f_s - 4(Dd\alpha^2 +2cT_1^2p_F^2)f_\sigma = 0,\notag\\
&D\partial_z f_t - 8\alpha^2 Ddf_t - c(T_0^2+4T_1^2p_F^2)f_t = 0,\notag\\
&D\partial_zf_s - 8c T_1^2p_F^2 f_s - 4i DT_1T_0 s^2Jm(f_\uparrow-f_\downarrow) \notag\\
&\;\;- 4cT_0^2f_s + 4s\e{i Jy_0}(T_0^2 + 4T_1^2p_F^2) = 0.
\end{align}
Here, $y_0$ is the position along the $y$-axis where the normal metal is coupled to the superconductor. 

From these boundary conditions, one obtains the solution for the unknown coefficients $A_{s,t,\sigma}$ and $B_{s,t,\sigma}$. With the anomalous Green function in hand, one may compute the spin magnetization induced in the normal metal:
\begin{align}
\vecM = M_0\int^\infty_{-\infty} \frac{d\varepsilon}{\varepsilon_\text{Th}} \text{Re} \text{Tr}\{\vecsigma\hat{g}^K\}
\end{align}
where $\varepsilon_\text{Th} = D/L^2$ is the Thouless energy and $M_0$ is a normalization constant of the same dimension as $\vecM$. 

The solution for the coefficients $A_{s,t,\sigma}$ and $B_{s,t,\sigma}$ are given in the appendix, and show that the anomalous Green function $f_t$ is zero whereas $f_\uparrow = -f_\downarrow$. Therefore, the only component of the induced magnetization that exists is $M_x$ which is determined by the product of the singlet anomalous Green function $f_s$ and the $x$-component of the triplet $d$-vector \cite{maeno_rmp_03}, $(f_\downarrow-f_\uparrow)/2$. From the expressions in the appendix, it then follows that
\begin{align}
M_x \propto JT_0T_1\Delta^2.
\end{align}
Therefore, the induced magnetization only exists in the presence of a supercurrent $(J)$ and in the presence of both normal tunneling $(T_0)$ and a Rashba-like tunneling $(T_1)$. We note that spin-orbit impurity scattering \cite{bergeret_prb_16} can cause a similar effect due to supercurrent flow. Moreover, we note that the effect predicted here is different from Ref. \cite{silaev_prb_20} since no exchange field is required anywhere in the junction.

\section{Conclusion}
In summary, we have derived a set of quasiclassical boundary conditions for spin-orbit coupled interfaces. The boundary conditions contain both terms related to transmission and reflection of particles and are valid both in and out of equilibrium. As an application, we have shown that a supercurrent flowing in a superconductor that is coupled to a normal metal through a Rashba-interface will induce an non-local magnetization in the normal metal. The magnetization is polarized in the direction perpendicular to both the supercurrent flow and the interface normal. These boundary conditions may prove useful to predict new spin-Hall like phenomena in superconducting hybrid structures and to model experimental data. 

\begin{acknowledgments}
E. H. Fyhn and S. Aunsmo is thanked for useful discussions. This work was supported by the Research Council of Norway through grant 240806, and its Centres of Excellence funding scheme grant 262633 ``\emph{QuSpin}''. 
\end{acknowledgments}

\begin{widetext}
\section{Appendix}

The coefficients determining the anomalous Green functions are obtained as
\begin{align}
B_{s,t,\sigma} = A_{s,t,\sigma} \e{2i kL}
\end{align}
and $A_\downarrow = -A_\uparrow$ where $e_\pm \equiv \e{\pm i k L}$ and
\begin{align}
A_\uparrow &= \frac{4i T_1DT_0mJ\e{i Jy_0}s\Big[ e_+\Big((-4T_0^2-16T_1^2p_F^2)c^2 + 4s^2(4T_1^2p_F^2+T_0^2)\Big) + i Dce_-k\Big]  }{\Big(-(T_0^2+4T_1^2p_F^2)^2c^2 - 2D\alpha^2 d(T_0^2+4T_1^2p_F^2)c - 8D^2m^2T_0^2T_1^2p_F^2J^2s^4\Big)8e_+^2 + 4i\Big((3T_0^2/2 + 6T_1^2p_F^2)c + D\alpha^2d\Big)ke_-De_+ + D^2e_-^2k^2}\notag\\
A_s &= \frac{4\e{i Jy_0} s\Big[ \Big(-16D^2J^2T_0^2T_1^2 cs^2m^2 - 4\alpha^2 d (4T_1^2p_F^2+T_0^2)D - 2c(T_0^2+4T_1^2p_F^2)(4T_1^2p_F^2 + T_0)\Big)e_+ + i(4T_1^2p_F^2 + T_0^2)kDe_-\Big]}{e_+^2\Big( -8(T_0^2 +4T_1^2p_F^2)^2c^2 - 16D\alpha^2 d(T_0^2+4T_1^2p_F^2)c - 64D^2m^2T_0^2T_1^2 J^2s^4\Big) + 4i kDe_+e_-\Big(D\alpha^2 d + 3c(T_0^2 +4T_1^2p_F^2)/2\Big) + D^2e_-^2k^2}
\end{align}
Note that the value of the interface location along the $y$-axis (the coordinate $y_0$) does not affect the value of physical observables such as the induced non-local magnetization.
 
\end{widetext}

\end{document}